\documentclass[oneside,english]{elsart}
\usepackage[T1]{fontenc}
\usepackage[latin1]{inputenc}
\usepackage{graphicx}
\usepackage{amssymb}

\makeatletter

\providecommand{\tabularnewline}{\\}

\usepackage{graphicx}

\usepackage{amssymb}
\usepackage{amsmath}

\usepackage{babel}
\makeatother
\begin{document}
\begin{frontmatter}

\title{A basis-set based Fortran program to solve the Gross-Pitaevskii Equation
for dilute Bose gases in harmonic and anharmonic traps}

\author{Rakesh Prabhat Tiwari$^{1,2}$, Alok Shukla$^{3}$}

\address{Physics Department, Indian Institute of Technology, Powai, Mumbai
400076, INDIA}

\thanks{Done in partial fulfillment of the requirements for the degree of
Bachelor of Technology at the Indian Institute of Technology, Bombay. }

\thanks{Present address: Department of Physics, The Ohio State University,
Columbus, OH 43210, USA. email:tiwari.12@osu.edu}

\thanks{Author to whom all the correspondence should be addressed. email:shukla@phy.iitb.ac.in}

\begin{abstract}
Inhomogeneous boson systems, such as the dilute gases of integral
spin atoms in low-temperature magnetic traps, are believed to be well
described by the Gross-Pitaevskii equation (GPE). GPE is a nonlinear
Schr\"odinger equation which describes the order parameter of such
systems at the mean field level. In the present work, we describe
a Fortran 90 computer program developed by us, which solves the GPE
using a basis set expansion technique. In this technique, the condensate
wave function (order parameter) is expanded in terms of the solutions
of the simple-harmonic oscillator (SHO) characterizing the atomic
trap. Additionally, the same approach is also used to solve the problems
in which the trap is weakly anharmonic, and the anharmonic potential
can be expressed in a polynomial in the position operators $x,$ $y$,
and $z$. The resulting eigenvalue problem is solved iteratively using
either the self-consistent-field (SCF) approach, or the imaginary
time steepest-descent (SD) approach. Iterations can be initiated using
either the simple-harmonic-oscillator ground state solution, or the
Thomas-Fermi (TF) solution. It is found that for condensates containing
up to a few hundred atoms, both approaches lead to rapid convergence.
However, in the strong interaction limit of condensates containing
thousands of atoms, it is the SD approach coupled with the TF starting
orbitals, which leads to quick convergence. Our results for harmonic
traps are also compared with those published by other authors using
different numerical approaches, and excellent agreement is obtained.
\textbf{}GPE is also solved for a few anharmonic potentials, and the
influence of anharmonicity on the condensate is discussed. Additionally,
the notion of Shannon entropy for the condensate wave function is
defined and studied as a function of the number of particles in the
trap. It is demonstrated numerically that the entropy increases with
the particle number in a monotonic way. \textbf{}
\end{abstract}
\end{frontmatter}
\textbf{Program Summary} \emph{}\\
\emph{Title of program:} bose.x \\
\emph{Catalogue Identifier:} \emph{}\\
\emph{Program summary URL:} \emph{}\\
\emph{Program obtainable from:} CPC Program Library, Queen's University
of Belfast, N. Ireland \\
\emph{Distribution format:} tar.gz\\
\emph{Computers :} PC's/Linux, Sun Ultra 10/Solaris, HP Alpha/Tru64,
IBM/AIX\\
\emph{Programming language used:} mostly Fortran 90\\
\emph{Number of bytes in distributed program, including test data,
etc.:} size of the tar file 153600 bytes\emph{}\\
\emph{Number of lines in distributed program, including test data,
etc.:} lines in the tar file 4\emph{221}\\
\emph{Card punching code:} ASCII\\
\emph{Nature of physical problem:} It is widely believed that the
static properties of dilute Bose condensates, as obtained in atomic
traps, can be described to a fairly good accuracy by the time-independent
Gross-Pitaevskii equation. This program presents an efficient approach
of solving this equation.\\
\emph{Method of Solution:} The solutions of the Gross-Pitaevskii equation
corresponding to the condensates in atomic traps are expanded as linear
combinations of simple-harmonic oscillator eigenfunctions. Thus, the
Gross-Pitaevskii equation which is a second-order nonlinear differential
equation, is transformed into a matrix eigenvalue problem. Thereby,
its solutions are obtained in a self-consistent manner, using methods
of computational linear algebra.\\
\emph{Unusual features of the program:} None

\section{Introduction}

Ever since the discovery of Bose-Einstein condensation (BEC) in dilute
atomic gases\cite{exp-1,exp-2,exp-3}, theoretical studies of this
and related phenomenon in such systems have grown exponentially\cite{stringari-rmp}.
For most of the theoretical studies of BEC in dilute gases, the starting
point is the so-called Gross-Pitaevskii equation (GPE)\cite{gross,pita},
which is nothing but a mean-field Schr\"odinger equation for a system
of Bosons interacting through a two-body interaction described by
$\delta$-function. In all but the simplest of the cases, one needs
to solve the GPE using numerical methods. For problems involving the
static properties of the condensate, the numerical solutions of the
time-independent GPE are of interest. And, indeed, over last several
years, a significant amount of work has been performed towards developing
novel approaches and algorithms meant for solving both time-dependent
and independent GPE. Next we survey some of the recent literature
in the field, restricting ourselves to the methods aimed at solving
the time-independent GPE, which is the subject of the present paper.
Edwards and Burnett developed a Runge-Kutta method based finite difference
approach for solving the time-independent GPE for spherical condensates\cite{edwards}.
In another paper Edwards \emph{et al.} used the basis set approach
similar to the one presented here, to solve the GPE for anisotropic
traps\cite{edwards-2}. Dalfovo and Stringari developed a finite-difference
based method for solving the time-independent GPE both for the ground
state, and the vortex states, in anisotropic traps\cite{dalfovo}.
Esry used a finite-element approach to solve for the both the time-independent
GPE, as well as, the Hartree-Fock equations for bosons confined in
anisotropic traps\cite{esry}. Schneider and Feder used a discrete
variable representation (DVR), coupled with a Gaussian quadrature
integration scheme, to obtain the ground and the excited states of
GPE in three dimensions\cite{schneider}. Adhikari used a finite-difference
based approach to solve the two-dimensional time-independent GPE\cite{adhikari-1,adhikari-2}.
Tosi and coworkers developed finite-difference, and imaginary-time,
approach for solving the time-independent GPE\cite{chiofalo}. Recently,
Bao and Tang developed a novel scheme for obtaining the ground state
of the GPE, by directly minimizing the corresponding energy functional\cite{bao1}.
Additionally, utilizing harmonic oscillator basis set, and Gauss-Hermite
quadrature integration scheme, Dion and Canc\`es have proposed a
scheme for solving both the time-dependent and -independent GPE\cite{dion}.
Earlier, we had also proposed an alternative scheme for dealing with
condensates with a large number of particles, and high number densities\cite{shukla}.

In this paper, we describe a Fortran program developed by us which
solves time-independent GPE corresponding to bosons trapped in Harmonic
traps. Instead of using the more common finite-difference approach,
we have chosen the basis-set based approach popular in quantum chemistry\cite{qchem}.
The basis set chosen for this case is the Cartesian simple-harmonic-oscillator
(SHO) basis set. The choice of a Cartesian basis set allows us to
treat the cases ranging from spherical condensates to completely anisotropic
condensates on an equal footing. Additionally, using the same approach
our program allows to solve the time-independent GPE for anharmonic
traps as well provided the anharmonic term can be expressed as a polynomial
in various powers of coordinates $x,\: y,$ and $z$. As far as the
SCF solution of the GPE is concerned, our program allows both the
matrix-diagonalization based scheme, as well as the use of the imaginary-time
steepest-descent method. Our program also allows the user to initiate
the SCF process using either the SHO ground state orbital, or the
Thomas-Fermi solution. We present the results of the calculations
performed with our code for several interesting cases, and very good
agreement is obtained with the existing results in the literature.
Additionally, in the present paper we have defined the notion of Shannon
entropy in the context of GPE, and presented various quantitative
calculations of the quantity.

Remainder of the paper is organized as follows. In the next section
we discuss the basic theoretical aspects of our approach. Next, in
section \ref{sec-program}, we briefly describe the most important
subroutines that comprise our program. In section \ref{sec-install}
we provide detailed explanation about installing and compiling our
program. Additionally, in the same section we explain how to prepare
an input file, and describe the contents of a typical output file.
In section \ref{sec-conv} we discuss the convergence properties of
the program with respect to the: (a) size of the basis set, and (b)
the method of solution. In section \ref{sec-example}, we discuss
results of several example runs of our program, and compare them to
those published earlier. Additionally, in the same section, we present
our results on the Shannon entropy of the condensate, and on the solutions
obtained in the presence of various anharmonic potentials. Finally,
in section \ref{sec-conclusions}, we present our conclusions. In
the Appendix we present the derivation of an analytical formula which
we have used in our program to compute the two-particle interaction
integrals.

\section{Theory}

\label{sec-theory}

For the present case, the time-independent Gross-Pitaevskii equation
is \begin{equation}
(-\frac{\hbar^{2}}{2m}\nabla^{2}+V_{\mbox{ext}}({\bf r})+\frac{4\pi\hbar^{2}aN}{m}|\psi({\bf r})|^{2})\psi({\bf r})=\mu\psi({\bf r}),\label{eq-gpe}\end{equation}
 where $\psi({\bf r})$ is condensate wave function one is solving
for, $V_{\mbox{ext}}({\bf r})=\frac{1}{2}m(\omega_{x}^{2}x^{2}+\omega_{y}^{2}y^{2}+\omega_{z}^{2}z^{2})+V^{anh}(x,y,z)$
is the confining potential for a general anisotropic trap ($\omega_{i}$'s
are the trap frequencies) with the anharmonic term $V^{anh}(x,y,z)$,
$a$ is the s-wave scattering length characterizing the two-body interactions
among the atoms, $N$ is the total number of bosons in the condensate,
and $\mu$ is the chemical potential. We are assuming that the condensate
wave function is normalized to unity. Before attempting numerical
solutions of Eq. (\ref{eq-gpe}), we cast it in a dimensionless form
by making the transformations\cite{bao1}

\begin{subequations}

\label{all-eq}

\begin{equation} \tilde{{\bf r}}=\frac{{\bf r}}{a_x}  \label{eq-all1} \end{equation} 

\begin{equation} \tilde{\psi}(\tilde{{\bf r}})=a_{x}^{3/2} \psi({\bf r}) \label{eq-all2}\end{equation} 

\

\end{subequations}

where $a_{x}=\sqrt{\frac{\hbar}{m\omega_{x}}}$ is the ``harmonic
oscillator length'' in the $x$ direction. This finally leads to the
dimensionless form of the GPE

\begin{equation}
(-\frac{1}{2}\tilde{{\nabla}^{2}}+V_{\mbox{ext}}(\tilde{{\bf r}})+\kappa|\tilde{\psi}(\tilde{{\bf r}})|^{2})\tilde{\psi}(\tilde{{\bf r}})=\tilde{\mu}\tilde{\psi}(\tilde{{\bf r}}),\label{eq-gpendim}\end{equation}

where $\tilde{\mu}$ is the chemical potential in the units of $\hbar\omega_{x}$,
$\kappa=\frac{4\pi Na}{a_{x}}$ is a dimensionless constant determining
the strength of the two-body interactions in the gas, and for the
harmonic oscillator potential $V_{\mbox{ext}}(\tilde{{\bf r}})=\frac{1}{2}(\tilde{x}^{2}+\gamma_{y}^{2}\tilde{y}^{2}+\gamma_{z}^{2}\tilde{z}^{2})+V^{anh}(\tilde{x},\tilde{y},\tilde{z})$,
with $\gamma_{y}=\frac{\omega_{y}}{\omega_{x}}$ and $\gamma_{z}=\frac{\omega_{z}}{\omega_{x}}$
being the two aspect ratios. We will now discuss the basis-set expansion
technique, used for solving the GPE\cite{edwards-2,schneider,dion}.
In this approach, one expands $\tilde{\psi}(\tilde{{\bf r}})$ as
a linear combination of basis functions of three-dimensional anisotropic
simple harmonic oscillator\begin{equation}
\tilde{\psi}(\tilde{{\bf r}})=\sum_{i=1}^{N_{basis}}C_{i}\Phi_{i}(\tilde{x},\tilde{y},\tilde{z})=\sum_{i=1}^{N_{basis}}C_{i}\phi_{n_{xi}}(\tilde{x})\phi_{n_{yi}}(\tilde{y})\phi_{n_{zi}}(\tilde{z}),\label{eq-basis}\end{equation}
 where $\phi_{n_{xi}}(\tilde{x})$, $\phi_{n_{yi}}(\tilde{y}$), and
$\phi_{n_{zi}}(\tilde{z})$, are the harmonic oscillator basis functions
corresponding to $x$, $y$, and $z$ directions, respectively, $C_{i}$
is the expansion coefficient, and $N_{basis}$ is the total number
of basis functions used. In dimensionless units, \emph{e.g.}, $\phi_{n_{zi}}(\tilde{z})$
can be written as\begin{equation}
\phi_{n_{zi}}(\tilde{z})=\left(\frac{\gamma_{z}}{\pi}\right)^{1/4}\frac{1}{\sqrt{2^{n_{zi}}n_{zi}!}}H_{n_{zi}}(\tilde{z}\sqrt{\gamma_{z}})\exp(-\frac{\gamma_{z}\tilde{z}^{2}}{2}),\label{eq-bfunc}\end{equation}
 where $H_{n_{zi}}(\tilde{z}\sqrt{\gamma_{z}})$ is a Hermite polynomial
of order $n_{zi}$ in the variable $\tilde{z}\sqrt{\gamma_{z}}$.
The form of the basis functions $\phi_{n_{xi}}(\tilde{x})$ and $\phi_{n_{yi}}(\tilde{y})$
can be easily deduced from Eq. (\ref{eq-bfunc}). Upon substituting
Eq. (\ref{eq-basis}) in Eq. (\ref{eq-gpendim}), then multiplying
both sides with another basis function $\Phi_{j}(\tilde{x},\tilde{y},\tilde{z})$
and integrating with respect to $\tilde{x}$, $\tilde{y}$, and $\tilde{z}$
the time-independent GPE is converted into an eigenvalue problem\cite{qchem}
\begin{equation}
\hat{F}\hat{C}=\tilde{\mu}\hat{C},\label{eq-eigval}\end{equation}
 where $\hat{C}$ represents the column vector containing expansion
coefficients $C_{i}$'s as its components, and the elements of the
matrix $\hat{F}$ are given by\begin{equation}
\hat{F_{i,j}}=E_{i}\delta_{i,j}+V_{i,j}^{anh}+g\sum_{k,l=1}^{N_{basis}}\tilde{J}_{i,j,k,l}C_{k}C_{l}.\label{eq-fock}\end{equation}
Above\begin{equation}
E_{i}=(n_{xi}+\frac{1}{2})+(n_{yi}+\frac{1}{2})\gamma_{y}+(n_{zi}+\frac{1}{2})\gamma_{z},\label{eq-ediag}\end{equation}
 $V_{i,j}^{anh}$ are the matrix elements of the anharmonic term in
the confining potential, and $\tilde{J}_{i,j,k,l}$ is the boson-boson
repulsion matrix. For anharmonic potentials which can be written as
polynomials in $x$, $y$, and $z$, $V_{i,j}^{anh}$ can be computed
quite easily, while $\tilde{J}_{i,j,k,l}$ can be written as a product
of three submatrices corresponding to the three Cartesian directions\cite{edwards-2}\begin{equation}
\tilde{J}_{i,j,k,l}=J_{n_{xi}n_{xj}n_{xk}n_{xl}}J_{n_{yi}n_{yj}n_{yk}n_{yl}}J_{n_{zi}n_{zj}n_{zk}n_{zl}.}\label{eq-jtilde}\end{equation}
 It can be shown that the elements of submatrices $J$ can be written
in the form\begin{equation}
J_{n_{i}n_{j}n_{k}n_{l}}=\int_{-\infty}^{\infty}d\xi\phi_{n_{l}}(\xi)\phi_{n_{k}}(\xi)\phi_{n_{j}}(\xi)\phi_{n_{i}}(\xi)\label{eq-jmat}\end{equation}
 where $\phi_{n_{i}}(\xi)$'s are the harmonic oscillator basis functions
of Eq. (\ref{eq-bfunc}). Integrals involved in Eq. (\ref{eq-jmat})
can be computed numerically using the methods of Gaussian quadrature\cite{schneider,dion},
or analytically\cite{edwards-2} using the formulas derived by Busbridge\cite{busbridge}.
In the present work, we have used this analytical expression---derived
in the appendix for the sake of completeness---to compute the values
of $J$ integrals.

The eigenvalue problem of Eq.(\ref{eq-eigval}) has to be solved selfconsistently.
In our program, for fixed values of $N$, this equation can be solved
for $\tilde{\mu}$ and $\hat{C}$ using either the iterative diagonalization
common in quantum chemistry\cite{qchem}, or the steepest-descent
approach as used by Dalfovo and Stringari\cite{dalfovo}. In both
the approaches, the onset of self-consistency is signalled once the
energy per particle of the condensate converges to within a user-specified
threshold. This approach is different from that of Edwards \emph{et
al.}\cite{edwards-2} where they solved Eq. (\ref{eq-eigval}) for
$N$, using fixed values of $\tilde{\mu}$. 

For the case of relatively small particle number, \emph{i.e.}, for
a weakly interacting system, it does not matter what is the nature
of starting guess for the condensate orbital for initiating the SCF
cycles. However, for the case of systems with large particle number,
the convergence obtained is very slow (if at all), in case the starting
condensate is taken to be the ground state of the harmonic trap. In
such cases, the convergence is easily obtained if the starting guess
for the condensate is taken to be of the Thomas-Fermi form, obtained
by setting the kinetic energy term in Eq. \ref{eq-gpendim} to zero\begin{equation}
|\tilde{\psi}_{TF}(\tilde{{\bf r}})|^{2}=\frac{\tilde{\mu}_{TF}-\frac{1}{2}(\tilde{x}^{2}+\gamma_{y}^{2}\tilde{y}^{2}+\gamma_{z}^{2}\tilde{z}^{2})}{g},\label{eq-tf}\end{equation}
where the Thomas-Fermi chemical potential $\tilde{\mu}_{TF}$ is given
by\begin{equation}
\tilde{\mu}_{TF}=\frac{\hbar\omega_{x}}{2}\left(\frac{15Na\gamma_{y}\gamma_{z}}{a_{x}}\right)^{2/5}.\label{eq-mutf}\end{equation}

In our program, we can also compute the Shannon entropy associated
with the condensate. The Shannon mixing entropy, for a general ensemble,
is defined as\cite{shannon}\begin{equation}
S=-\sum_{i}P_{i}\log\: P_{i}\;,\label{eq-shannon}\end{equation}
where $P_{i}$ is the probability for a system to be in the $i$-th
state. Of course, in the present case, we do not have a thermodynamic
system in which the mixing of various states will take place due to
thermal fluctuations driven by its finite temperature. Thus, the question
is as to how to define Shannon entropy for the present system, which
is essentially being treated as a zero-temperature quantum mechanical
system. For the purpose we adopt an information theoretic point-of-view,
and define the probability $P_{i}$ as\begin{equation}
P_{i}=|C_{i}|^{2}.\label{eq-pi}\end{equation}
where $C_{i}$ are the expansion coefficients of various SHO eigenstates
in the expression for the condensate wave function of Eq. (\ref{eq-basis}).
In this picture, ground state of the condensate is seen as a statistical
mixture of the various eigenstates of SHO, with the mixing probability
$P_{i}$. It is important to realize that the reason behind this mixing
of states is the inter-particle interaction in the condensate, because,
in its absence, the condensate will be in the ground state of the
SHO ($C_{i}=\delta_{1,i}$) leading to $S=0$. Thus, in a sense, entropy
defined as per Eqs. (\ref{eq-shannon}) and (\ref{eq-pi}) is a measure
of inter-particle interactions in the system. Because, stronger the
inter-particle interactions, the condensate will be a mixture of larger
number of states, leading to a larger entropy. From an information-theoretic
point-of-view larger entropy implies loss of information about the
system, because, in such a case, the system is a mixture of a larger
number of states. The point to be remembered, however, is that this
information loss is being driven by the inter-particle interactions
in the system while the corresponding information loss in a thermodynamic
system is driven by its finite temperature, and the thermal fluctuations
caused by it. In various contexts, other authors have also computed
and discussed the information entropy associated with interacting
quantum systems\cite{inf-gadre,inf-ziesche,inf-moust}.

\section{Description of the program}

\label{sec-program}

In this section we briefly describe the main program and various subroutines
which constitute the entire module. All the subroutines, except for
the diagonalization subroutine taken from EISPACK\cite{househ}, have
been written in the Fortran 90 language.

\subsection{Main Program OSCL}

The main program is called OSCL, and its task is to read all the input
information, and, among other things, fix the dimensions of various
arrays. All the arrays needed in the program are dynamically allocated
either in the main program, or in some of the subroutines. By utilizing
the dynamic array allocation facility of the Fortran 90 language,
we have made the program independent of the size of the calculations
undertaken. Because of this, the program needs to be compiled only
once, and will run until the point the memory available on the computer
is exhausted. Besides reading all the necessary input, program OSCL
calls various subroutines in which different tasks associated with
the calculation are performed.

\subsection{XMAT\_0 }

XMAT\_0 is a small subroutine whose job is to compute the matrix elements
of the position operator in harmonic oscillator units, with respect
to the basis set of a one-dimensional SHO. This routine is called
from the main program OSCL, and results are stored in a two-dimensional
array called $xmatrx$.

\subsection{Basis Set Generation}

The calculations are performed using a basis set of a three-dimensional
SHO consistent with the symmetry of the system. The basis set to be
used is generated by calling one of the following three routines:
(a) for a spherical condensate (complete isotropy) routine BASGEN3D\_ISO
is used, (b) for a cylindrical condensate routine BASGEN3D\_CYL is
called, and (c) the basis set for a completely anisotropic condensate
is generated using the routine BASGEN3D\_ANISO. In all the cases the
basis functions are arranged in the ascending order of their harmonic
oscillator energies and, if needed, a heap sort routine called HPSORT
is used to achieve that. All these subroutines have the option of
imposing parity symmetry on the basis set if the potential has that
symmetry. This leads to a substantial reduction in the size of the
basis set in most cases.

\subsection{HAM0\_3D:}

This subroutine is called from the main program OSCL and its purpose
is to generate the matrix elements of the noninteracting (one-particle)
part of the condensate Hamiltonian. If the condensate is confined
in a perfectly harmonic trap, one-particle part of the Hamiltonian
is trivial. However, for the case where the trap potential is anharmonic,
the potential matrix elements are generated from the position operator
matrix elements $xmatrx(i,j)$ mentioned above.

\subsection{BEC\_DRV}

Subroutine BEC\_DRV is called from the main program OSCL, and as its
name suggests, it is the driver routine for performing calculations
of the condensate using the time-independent GPE. Apart from allocating
a few arrays, the main task of this routine is to call either: (a)
routine BOSE\_SCF meant for solving for the condensate wave function
using the iterative-diagonalization-based SCF approach, or (b) routine
BOSE\_STEEP used for solving for the condensate wave function using
the steepest-descent approach of Dalfovo and Stringari\cite{dalfovo}.

\subsection{BOSE\_SCF}

This subroutine solves the time-independent GPE in a self-consistent
manner using an iterative diagonalization approach. Its main tasks
are as follows:

\begin{enumerate}
\item Allocate various arrays needed for the SCF calculations
\item Setup the starting orbitals. For this the options are: (i) diagonalize
the one-particle part of the Hamiltonian, (ii) use the Thomas-Fermi
orbitals, or (iii) use the orbitals obtained in a previous run.
\item Perform the SCF calculations. For the purpose, the two-particle integrals
$\tilde{J}_{i,j,k,l}$ (cf. Eq. (\ref{eq-jtilde}) are calculated
on the fly during each iteration using the formulas derived in the
appendix. In other words the storage of these matrix elements is completely
avoided, thereby saving substantial amount of memory and disk space.
This approach is akin to the ``direct SCF'' approach utilized in
quantum chemistry. Permutation symmetries of indices $i,\: j,\: k,$
and $l$ are utilized to reduce the number of integrals evaluated.
Moreover, the evaluation of an integral is undertaken only if it is
found to be nonzero as per the symmetry selection rules. The integrals
in question are evaluated in a subroutine called JMNPQ\_CAL. The $\hat{F}$
matrix constructed in each iteration is diagonalized through a Householder
diagonalization routine called HOUSEH, which is from the EISPACK package
of routines\cite{househ}, and is written in Fortran 77.
\item The chemical potential and the condensate wave function obtained after
every iteration are written in various data files so that the progress
of the calculation can be monitored.
\end{enumerate}

\subsection{BOSE\_STEEP}

Alternatively, the condensate wave function and the chemical potential
can be obtained using the subroutine BOSE\_STEEP which, instead of
the iterative diagonalization approach, utilizes the steepest-descent
approach to achieve convergence, starting from a given starting orbital.
In this approach, as outlined by Dalfovo and Stringari\cite{dalfovo},
the starting orbitals are evolved towards the true orbitals in small
imaginary time steps by repeated application of the Hamiltonian, i.e.,
the $\hat{F}$ operator. Here, the main computational step is the
multiplication of a column vector by a matrix, which in the present
version of the program is achieved by a call to the Fortran 90 intrinsic
function MATMUL. However, one could certainly try to improve upon
this by developing a subroutine which can utilize the symmetric nature
of the Fock matrix. Apart from this, rest of the actions performed
in this subroutine are identical to those of BOSE\_SCF.

\subsection{THOMAS\_FERMI}

This subroutine is invoked either from the subroutine BOSE\_SCF or
from BOSE\_STEEP in case the SCF calculations are to be initiated
by assuming Thomas-Fermi form of the starting orbitals. Upon invocation,
this subroutine directly constructs the operator $\hat{F}$ corresponding
to the Thomas-Fermi orbitals. In this case, the ${\bf r}$-space integration
is performed using a trapezoidal-rule-based scheme on a three-dimensional
Cartesian grid.

\subsection{ENTROPY}

This subroutine is called if the entropy of the condensate needs to
be computed with respect to the Harmonic oscillator basis functions,
as per Eqs. \ref{eq-shannon} and \ref{eq-pi}. It is a very small
subroutine with a straightforward implementation.

\subsection{COND\_PLOT}

This subroutine computes the numerical values of the condensate wave
function for a user-specified set of points in space. The numerical
values of the Hermite polynomials needed for the purpose are computed
using the subroutine HERMITE, described below.

\subsection{HERMITE}

This subroutine computes the values of the Hermite polynomial $H_{n}(x)$
for a set of user specified values of $x$, and order $n$. Fast computation
of polynomials is achieved by using initializations $H_{0}(x)=1$,
$H_{1}(x)=2x$, and the recursion relation $H_{n+1}(x)=2xH_{n}(x)-2nH_{n-1}(x)$.

\section{Installation}

\label{sec-install}

All the files needed to install and run the program are kept in the
gzipped, tarred archive \texttt{bose.tar.gz}. It consists of: (a)
All the Fortran files containing the main program (file \texttt{oscl.f90}),
and various subroutines, called by the main program, (b) four versions
of Makefiles which can be used for compiling the code on various Linux/Unix
systems, and (c) several sample input and output files in a subdirectory
called \texttt{Examples}. The program was developed on Pentium 4 based
machine running Redhat Fedora core 1 operating system using noncommercial
version of the Intel Fortran compiler version 8.1. However, it has
also been verified that it runs on Sun Solaris Sparc based systems,
Compaq alpha (now HP alpha) based systems running True Unix, and IBM
Power PC systems running AIX. For these systems, the Fortran 90 compilers
supplied with those operating systems were used. In order to install
and compile the program, following steps need to be followed:

\begin{enumerate}
\item Uncompress the program files in a directory of user's choice using
the command \texttt{gunzip bose.tar.gz} followed by \texttt{tar -xvf
bose.tar.}
\item Verify that the four makefiles \texttt{Makefile\_linux}, \texttt{Makefile\_solaris},
\texttt{Makefile\_alpha,} and \texttt{Makefile\_aix} are present.
Copy the suitable version of the make file to the file \texttt{Makefile}.
For example, if the system is a Sun Solaris Sparc system, issue the
command \texttt{cp Makefile\_solaris Makefile}.
\item Now issue the command \texttt{make} which will initiate the compilation.
If everything is successful, upon completion \texttt{bin} directory
of your account will have the program execution file \texttt{bose.x}.
If your account does not have a directory named \texttt{bin}, you
will have to either create this directory, or modify the \texttt{Makefile}
to ensure that the file \texttt{bose.x} is created in the directory
of your choice.
\item If the \texttt{bin} directory is in your path, try running the program
using one of the sample input files located in the subdirectory \texttt{Examples}.
For example, by issuing the command \texttt{bose.x < bec\_iso.dat
> x.out} one can run the program for an isotropic trap and the output
will be written in a file called \texttt{x.out}. This should be compared
with the supplied file \texttt{bec\_iso.out} to make sure that results
obtained agree with those of the example run.
\end{enumerate}
Additionally, a file called \texttt{README} is also provided which
lists and briefly explains all the files included in the package.
Although we have not investigated the installation of the program
on operating systems other than Linux/Unix, we do not anticipate any
problems with such operating systems.

\subsection{Input Files}

\label{sub-input} In order to keep the input process as free of errors
as possible, we have adopted the philosophy that before each important
input cards, there will be a compulsory comment line. It is irrelevant
as to what is written in the comment lines, but, by writing something
meaningful, one can keep the input process transparent. The input
quantities following the comment line have to be in free format, with
the restriction that the ASCII input cards should be in uppercase
letters. Because of the use of comment lines, the input files are
more or less self explanatory. In the sample input files, we have
started all the comment lines with the character \texttt{\#}. A sample
input file corresponding to a cylindrical trap potential is listed
below

\texttt{\#Type of oscillator}

\texttt{CYLINDRICAL }

\texttt{\# NXMAX, NYMAX, NZMAX}

\texttt{10, 8 }

\texttt{\# NO. OF TERMS IN THE ANHARMONIC POTENTIAL}

\texttt{0}

\texttt{\# OMEGAX, OMEGAY, OMEGAZ, JILA parameters}

\texttt{1.0, 2.8284271}

\texttt{\# Type of SCF equation}

\texttt{GP }

\texttt{\# No. of particles}

\texttt{1000 }

\texttt{\# Scattering Length, JILA Rb87}

\texttt{4.33d-3 }

\texttt{\# SCF convergence threshold, Maximum \# of allowed iterations.}

\texttt{1.d-8, 1000}

\texttt{\# Whether Parity is a good quantum number or not}

\texttt{PARITY}

\texttt{\# Method for calculations}

\texttt{SCF}

\texttt{\# Starting orbitals}

\texttt{SHO }

\texttt{\# Whether orbital Mixing will be done}

\texttt{FOCKMIX}

\texttt{0.4}

\texttt{\# Whether orbital plots needed}

\texttt{PLOT}

\texttt{-5.0,5.0,0.05}

\texttt{1}

\texttt{1,0,0}

\texttt{\# Entropy Calculation}

\texttt{ENTROPY}

\texttt{1,1}

Next we describe the input cards one by one.

\begin{enumerate}
\item First card is an ASCII card describing the type of trap potential.
Options are: \texttt{ISOTROPIC}, \texttt{CYLINDRICAL}, or \texttt{ANISOTROPIC}.
\item Second card specifies the maximum quantum numbers of the basis functions
to be included for various directions. For an isotropic trap one entry
is needed ($n_{x}=n_{y}=n_{z}$), for cylindrical trap, two entries
are needed ($n_{x}=n_{y}$ and $n_{z}$), while for an anisotropic
oscillator three entries are needed ($n_{x}$, $n_{y}$, $n_{z}$).
These numbers eventually determine the total number of basis functions
$N_{basis}$ used to expand the condensate as per Eq. \ref{eq-basis}.
\item Third card deals with the anharmonic terms in the trap potential.
Any potential of the form $\sum_{i=1}^{N_{anh}}C_{i}x^{m_{i}^{x}}y^{m_{i}^{y}}z^{m_{i}^{z}}$
can be added to the Harmonic trap potential. The first entry here
is $N_{anh}$ after which $N_{anh}$ entries consisting of $\{ C_{i},m_{i}^{x},m_{i}^{y},m_{i}^{z}\}$
are given. In the example input, no anharmonicity was considered,
thus $N_{anh}$ has been set to zero.
\item Fourth card deals with the trap frequencies with the convention that
$\omega_{x}=1$, and rest of the frequencies measured in the units
of $\omega_{x}$. The example input corresponds to the trap frequencies
of the JILA experiment with $\omega_{x}=\omega_{y}=1,\;\omega_{z}=2.284271$.
\item Fifth card specifies which mean-field equation is to be solved. Options
are \texttt{GP} for the Gross-Pitaevskii equation, and \texttt{HF}
for the Hartree-Fock equation.
\item Sixth card reads the total number of bosons in the trap.
\item Seventh card is the value of the $s$-wave scattering length in the
Harmonic oscillator units.
\item Eighth card inputs the convergence threshold, followed by the maximum
number of iterations allowed to achieve convergence.
\item Ninth card specifies whether parity should be treated as a good quantum
number or not. Options are \texttt{PARITY} and \texttt{NOPARITY}.
If the trap potential is invariant under the parity operation, use
of this card leads to a tremendous reduction in the size of the basis
set needed to solve for the condensate wave function.
\item Tenth card specifies as to which method is to be used for solving
the mean-field equations. Options are \texttt{SCF} corresponding to
the iterative diagonalization method, and \texttt{STEEPEST-DESCENT}
corresponding to the use steepest-descent method of Dalfovo and Stringari\cite{dalfovo}.
In case one opts for the steepest-descent approach, the size of the
time step to be used in the calculations also needs to be specified.
\item Eleventh card specifies as to what sort of starting guess for the
condensate should be used to start the solution process. Valid options
are \texttt{SHO} corresponding to the simple-harmonic oscillator ground
state and \texttt{THOMAS-FERMI} corresponding to the Thomas-Fermi
form of the condensate. 
\item It has been found that in several difficult cases, convergence can
be achieved if one utilizes the techniques of Fock matrix mixing or
condensate orbital mixing\cite{dalfovo,esry}. Valid options are (a)
\texttt{FOCKMIX} for Fock matrix mixing (b) \texttt{ORBMIX} for condensate
mixing, (c) Any other ASCII entry such as \texttt{NOMIX} for neither
of these options. In case options (a) or (b) are chosen, one needs
to specify the parameter $xmix$ quantifying the mixing according
to the formula\[
R^{(i)}=xmix\: R^{(i)}+(1-xmix)\: R^{(i-1)},\]
where $R^{(i)}$ is the quantity under consideration in the $i$-th
iteration. Thus, if Fock matrix mixing has been opted, $xmix$ specifies
the fraction of the new Fock matrix in the total Fock matrix in the
$i$-th iteration.
\item This is the penultimate card which decides whether the user wants
the numerical values of condensate wave function along user specified
set of data points, such that the condensate could be plotted as a
function of spatial coordinates. Keyword PLOT means that the answer
is in affirmative while any other option such as NOPLOT will disable
the numerical computation of the condensate. If the keyword PLOT has
been supplied as in the example input, further data $rmin,\: rmax,\: dr$
is specified next which determines the starting position, ending position,
and the step size for generating the points on which the condensate
is to be computed. After these values, we need to specify variable
$ndir$ which is the number of directions along which the condensate
needs to be computed. Finally, $ndir$ Cartesian directions have to
be specified. The example input file instructs the program to compute
the value of the condensate along the $x$ axis, for $-5.0\leq x\leq5.0$,
in the steps of 0.05. 
\item Final card specifies whether one wants to compute the mixing entropy.
Valid options are \texttt{ENTROPY} and any other entry such as \texttt{NOENTROPY.}
If one opts for entropy calculation, one can do so for a whole range
of eigenfunctions specified by their lower bound and the upper bound.
Entry \texttt{1,1} specifies that entropy of only the ground state
needs to be computed.
\end{enumerate}

\subsection{Output file}

\label{sub-output}

Apart from the usual information related to various system parameters,
the most important information that an output file contains is the
approach (or lack thereof) to convergence of the calculations as far
as the chemical potential is concerned. Besides that, any other computed
quantity such as the entropy is also listed in the output file. The
important portions of the output file, corresponding to the input
file discussed in the previous section, are reproduced below. The
complete sample output file is called \texttt{bec\_cyl\_jila.out},
and is included in the tar archive.

\texttt{SCF iterations begin}

\texttt{Starting chemical potential= 2.4142135}

\texttt{Iteration \# \hspace*{1.0cm}Chem. Pot. \hspace*{1.0cm}Energy/particle
\hspace*{1.0cm}Energy-Converg.}

\texttt{\hspace*{1.0cm}1 \hspace*{2.2cm}3.876643\hspace*{2.0cm}5.3193762
\hspace*{2.2cm}5.3193762}

\texttt{\hspace*{1.0cm}2 \hspace*{2.2cm}4.271632 \hspace*{1.8cm}4.0689872
\hspace*{2.0cm}-1.2503890}

\texttt{\hspace*{1.0cm}3 \hspace*{2.2cm}4.517259 \hspace*{1.8cm}3.8580062
\hspace*{2.0cm}-0.2109810}

\texttt{\hspace*{1.0cm}4 \hspace*{2.2cm}4.619476 \hspace*{1.8cm}3.8439807
\hspace*{2.0cm}-0.0140255}

\texttt{\hspace*{1.0cm}5 \hspace*{2.2cm}4.689213 \hspace*{1.8cm}3.8418453
\hspace*{2.0cm}-0.0021354}

\texttt{\hspace*{1.0cm}6 \hspace*{2.2cm}4.720534 \hspace*{1.8cm}3.8413295
\hspace*{2.0cm}-0.0005159}

\texttt{\hspace*{1.0cm}7 \hspace*{2.2cm}4.743576 \hspace*{1.8cm}3.8411720
\hspace*{2.0cm}-0.0001575}

\texttt{\hspace*{1.0cm}8 \hspace*{2.2cm}4.753914 \hspace*{1.8cm}3.8411212
\hspace*{2.0cm}-0.0000508}

\texttt{\hspace*{1.0cm}9 \hspace*{2.2cm}4.761791 \hspace*{1.8cm}3.8411040
\hspace*{2.0cm}-0.0000172}

\texttt{\hspace*{1.0cm}10 \hspace*{2.1cm}4.765292 \hspace*{1.7cm}3.8410982
\hspace*{2.0cm}-0.0000059}

\texttt{\hspace*{1.0cm}11 \hspace*{2.1cm}4.768022 \hspace*{1.8cm}3.8410961
\hspace*{2.0cm}-0.0000020}

\texttt{\hspace*{1.0cm}12 \hspace*{2.1cm}4.769223 \hspace*{1.8cm}3.8410954
\hspace*{2.0cm}-0.0000007}

\texttt{\hspace*{1.0cm}13 \hspace*{2.1cm}4.770177 \hspace*{1.8cm}3.8410951
\hspace*{2.0cm}-0.0000003}

\texttt{\hspace*{1.0cm}14 \hspace*{2.1cm}4.770592 \hspace*{1.8cm}3.8410950
\hspace*{2.0cm}-0.0000001}

\texttt{\hspace*{1.0cm}15 \hspace*{2.1cm}4.770927 \hspace*{1.8cm}3.8410950
\hspace*{2.2cm}0.0000000}

\texttt{\hspace*{1.0cm}16 \hspace*{2.1cm}4.771071 \hspace*{1.8cm}3.8410950
\hspace*{2.2cm}0.0000000}

\texttt{\hspace*{1.0cm}17 \hspace*{2.1cm}4.771189 \hspace*{1.8cm}3.8410950
\hspace*{2.2cm}0.0000000}

\texttt{Convergence achieved on the BEC ground state}

\texttt{Eigenstate \# \hspace*{1.0cm}Information Entropy}

\texttt{\hspace{1.0cm}1 \hspace*{3.2cm}0.7477159}

The contents of the output file listed above are self-explanatory.
It basically shows that after seventeen iterations, the total energy
per particle of the condensate converges to the value 3.841095, leading
to the chemical potential value of 4.771. Additionally, the entropy
of the condensate is computed to be 0.744771.

In addition to the above mentioned main output file, there is another
output file created in the ASCII format which contains the condensate
orbital obtained at the end of each iteration. This file is written
in the logical unit 9, and is named \texttt{orbitals.dat}. When a
new run is started, the program always looks for this file and tries
to use the condensate solution present there to start the iterations.
In other words, condensate solution present in \texttt{orbitals.dat}
is used to restart an old aborted run. If some incompatibility is
found between the condensate solution, and the present run, the solution
in the \texttt{orbitals.dat} is ignored and a new run is initiated.
Thus, if one wants to start a completely new run, any old \texttt{orbitals.dat}
file must first be deleted.

\section{Convergence Issues}

\label{sec-conv}In this section we compare the convergence of our
results with respect to the size of the basis set used. We also compare
the convergence properties of different iterative approaches aimed
at obtaining the condensate solutions.

\subsection{Convergence with respect to the basis set}

\label{sub-basis}

Before treating the results obtained as the true results, one must
be sure as to convergence properties with respect to the size of the
basis set. This aspect of the calculations is explored in the present
section by means of two examples corresponding to condensates in spherical
and cylindrical traps, respectively, with fifteen hundred ($N=1500)$
bosons each. 

First we discuss the case of the condensate in an isotropic trap,
results for which are presented in table \ref{tab-conv-iso}. The
value of the scattering length used in the calculations is listed
in the caption of the table. For this case, only one value specifying
the largest quantum number \texttt{nmax} for the basis functions,
needs to be specified. It is obvious from the table that the results
which have converged to three decimal places both in the chemical
potential and the entropy require $nmax=8$, leading to the total
number of basis functions $N_{basis}=35$. This means that the size
of the Fock matrix diagonalized during the iterative diagonalization
is $35\times35$, which is computationally very inexpensive. It is
also obvious from the table that in order to get four decimal place
convergence, we only need to use $nmax=10$ corresponding to a $56\times56$
Fock matrix, whose diagonalization can also be carried out quite fast.

Similar results for the condensate in a cylindrical trap corresponding
to the JILA parameters\cite{jila} are presented in table \ref{tab-conv-cyl}.
Because of the anisotropy of the trap, the convergence is to be judged
with respect to two parameters $nxmax$ deciding the highest quantum
number of the basis functions for $x-$ and $y-$directions, and $nzmax$
the corresponding number for the $z$-direction. We will first try
to understand the convergence properties using a few heuristic arguments.
In the JILA experiment\cite{jila} the trap frequency in the $z$-direction
$\omega_{z}$ was more than twice the value of the trap frequencies
in the $x$- and $y$-directions, $\omega_{x}$ and $\omega_{y}$.
Therefore, due to inter-particle repulsion, the condensate will be
much more delocalized along the $x/y$-directions, as compared to
the $z$-direction. This means that, in order to achieve convergence,
one would expect to use higher-energy basis functions in the $x/y$-directions,
as compared to the $z$-direction. In other words, at convergence
$nxmax>nzmax$. And when we examine table \ref{tab-conv-cyl}, we
find that this is indeed the case. We notice that the three-decimal
place convergence in both the chemical potential, and the entropy,
is obtained for $nxmax=8$ and $nzmax=6$, although the data presented
in the table covers a much larger range of parameters. Thus, we conclude
that reasonably accurate values of various physical quantities can
be obtained with basis sets of modest sizes, both for the isotropic
as well as for the cylindrical condensates.

\subsection{Comparison of different numerical approaches}

\label{sub-solution}As mentioned earlier, our program can solve the
GP equation using two numerical approaches: (a) iterative diagonalization
(ID) of the Fock matrix, and (b) steepest-descent (SD) method of Dalfovo
\emph{et al}.\cite{dalfovo}. In the previous sections, all the presented
results were obtained by the ID method. In the present section, we
would like to present results based upon the steepest-descent approach,
and compare them to those obtained using the iterative diagonalization
method. We present our results for the cylindrical trap corresponding
to the JILA parameters\cite{jila}, with an increasing number of particles
in table \ref{tab-comp-method}. Obviously, the numerical solution
of the GP equation becomes increasingly difficult as the number of
particles in the condensate grows, because of the increased contribution
of the inter-particle repulsion. Therefore, it is very important to
know as to how various numerical approaches perform as $N$ is gradually
increased. 

As the results presented in the table suggest that for smaller values
of $N$, neither of the two approaches have any problems achieving
convergence, and the results obtained were found to be in agreement
with each other to three decimal places for the basis set used. We
found that in most of the cases, the ID approach worked only when
Fock-matrix mixing used. Although, we managed to achieve convergence
for the cases depicted in table \ref{tab-comp-method} with the ID
method; however, as the number of bosons in the condensate grows further,
the convergence becomes slow and difficult to achieve by this method,
a fact also emphasized by Schneider and Feder\cite{schneider}. On
the other hand, the SD approach did not have any convergence problems
for the cases we investigated. As depicted in table \ref{tab-comp-method},
the SD approach led to convergence both with the SHO starting orbital,
as well as the Thomas-Fermi starting orbital. However, for larger
values of $N$, the use of Thomas-Fermi solution as the starting guess
for the condensate, will lead to much faster convergence with this
approach. Thus, we conclude that: (a) For smaller values of $N$,
both the ID as well as the SD methods will lead to convergence, and
(b) for really large values of $N$, the convergence is guaranteed
only with the SD method. In the SD method the main computational operation
is the multiplication of a vector by a matrix, which will be significantly
faster as compared to the matrix diagonalization procedure needed
by the ID method for calculations involving large basis sets. Thus,
for calculations involving large basis sets, SD method will be faster
as compared to the ID method. Therefore, all things considered, we
believe that the SD method is the more robust of the two possible
approaches.

\section{Example Runs}

\label{sec-example}In this section we report the results of our calculations
for a variety of trap parameters, and compare our results with those
published by other authors. We also study the behavior of the Shannon
entropy of the condensate with respect to the number of particles
it contains. Additionally, we also present our results for the cases
of anharmonic traps.

\subsection{Comparison with other works}

\label{sub-comp}

In this section we present the results of several calculations performed
on both isotropic and the cylindrical traps, and compare them to the
results obtained by other authors. Several authors have performed
such calculations, however, for the sake of brevity, we restrict our
comparisons mainly to the works of Bao and Tang\cite{bao1} for the
spherical condensate, and to the results of Dalfovo and Stringari\cite{dalfovo}
for the cylindrical condensate.

Recently, using finite-element based approach, Bao and Tang\cite{bao1}
performed calculations for condensates on a variety of harmonic traps,
and presented results as a function of the interaction parameter $\kappa=\frac{4\pi aN}{a_{x}}$.
In table \ref{tab-bao} our results for the chemical potentials of
condensates in isotropic traps corresponding to increasing values
of $\kappa$ are compared with those reported by Bao and Tang\cite{bao1}.
The agreement between the results obtained by two approaches is exact
to the decimal places reported by Bao and Tang\cite{bao1}. Note that
the aforesaid agreement was obtained for rather modest basis set sizes,
and calculations were completed on a personal computer in a matter
of minutes. 

Next we discuss the results obtained for a cylindrical trap corresponding
to JILA parameters\cite{jila} for an increasing number of bosons.
Our results are presented in table \ref{tab-dalfovo}, where they
are also compared to the results of Dalfovo and Stringari obtained
using a finite-difference based approach\cite{dalfovo}. The agreement
between our results and those of Dalfovo and Stringari is virtually
exact for all their reported calculations\cite{dalfovo}. Again, the
noteworthy point is that this level of agreement was obtained with
the use of modest sized basis sets, and the computer time running
into a few minutes.

Thus, the excellent agreement between our results, and those obtained
by other authors using different approaches, gives us confidence about
the essential correctness of our methodology. Now the question arises,
will this numerical method work for values of the interaction parameter
$\kappa$ which are much larger than the ones considered here. The
encouraging aspect of the approach is that for none of the larger
values of $\kappa$ which we considered did we experience a numerical
breakdown of the approach. It is just that for larger values of $\kappa$,
the total number of basis functions needed to achieve convergence
on the chemical potential will be larger as compared to the smaller
$\kappa$ cases. This, of course, will also lead to an increase in
the CPU time needed to perform converged calculations. For example,
for the case of the spherical trap considered in table \ref{tab-bao},
when we doubled $\kappa$ to the value $6274$, we needed to use basis
functions corresponding to $nmax=20$ with $N_{basis}=286$ to achieve
two-decimal places convergence in the chemical potential. For $\kappa=9411$,
to achieve similar convergence, these numbers increased to $nmax=24$
and $N_{basis}=455$. Finally, when $\kappa$ was increased to $15685$,
the corresponding numbers were $nmax=28$ and $N_{basis}=680$, with
the CPU time running into several hours. For the cylindrical trap
(cf. table \ref{tab-dalfovo}), for $N=20000$ bosons ($\kappa=1088.2$)
the convergence was achieved in a matter of minutes with $nxmax=14$,
$nzmax=8$ and $N_{basis}=180$. When the number of bosons in the
trap was doubled such that $\kappa=2176.4$, similar level of convergence
on the chemical potential was obtained with $nxmax=18$, $nzmax=8$
and $N_{basis}=275$. Even with much larger values of $\kappa$ ($>30000$)
both for the spherical, and the cylindrical traps, we did not encounter
any convergence difficulties when the calculations were performed
with the modest sized basis sets mentioned earlier. But it was quite
obvious that, to obtain highly accurate values of chemical potentials
for such cases, one will have to use basis sets running into thousands
which will make the calculations quite time consuming.

At this point, we would also like to compare our approach to that
of Schneider and Feder\cite{schneider}, who used a DVR based technique
to obtain accurate solutions of the time-independent GPE. In the DVR
approach the basis functions are the so-called ``coordinate eigenfunctions'',
which, in turn, are assumed to be linear-combinations of other functions
such as the SHO eigenstates, or the Lagrange interpolating functions\cite{schneider}.
Thus in the DVR approach of Schneider and Feder\cite{schneider},
the SHO eigenstates are used as intermediate basis functions, and
not as primary basis functions as is done in our approach. Using this
approach, coupled with the ``direct-inversion in the iterative space''
(DIIS) method, Feder and Schneider managed to obtain accurate solutions
for anisotropic condensates for quite large values of the interaction
parameter $\kappa$\cite{schneider}. However, the price to be paid
for this accuracy was the use of a very large basis set consisting
of several thousands of basis functions\cite{schneider} even for
rather small values of $\kappa$.

\begin{figure}

\caption{Plots of condensates along the $x$ axis for an isotropic trap, with
an increasing number of $N$ of bosons. The trap parameters used were
the same as in the data of tables \ref{tab-conv-iso} and \ref{tab-bao}.
Lines correspond to $N=100,\:500,\:1000,\:5000,$ and 10000, and are
in the descending order of the central condensate density, and distances
are in the units of $a_{x}$. }

\includegraphics[%
  scale=0.8,
  angle=-90]{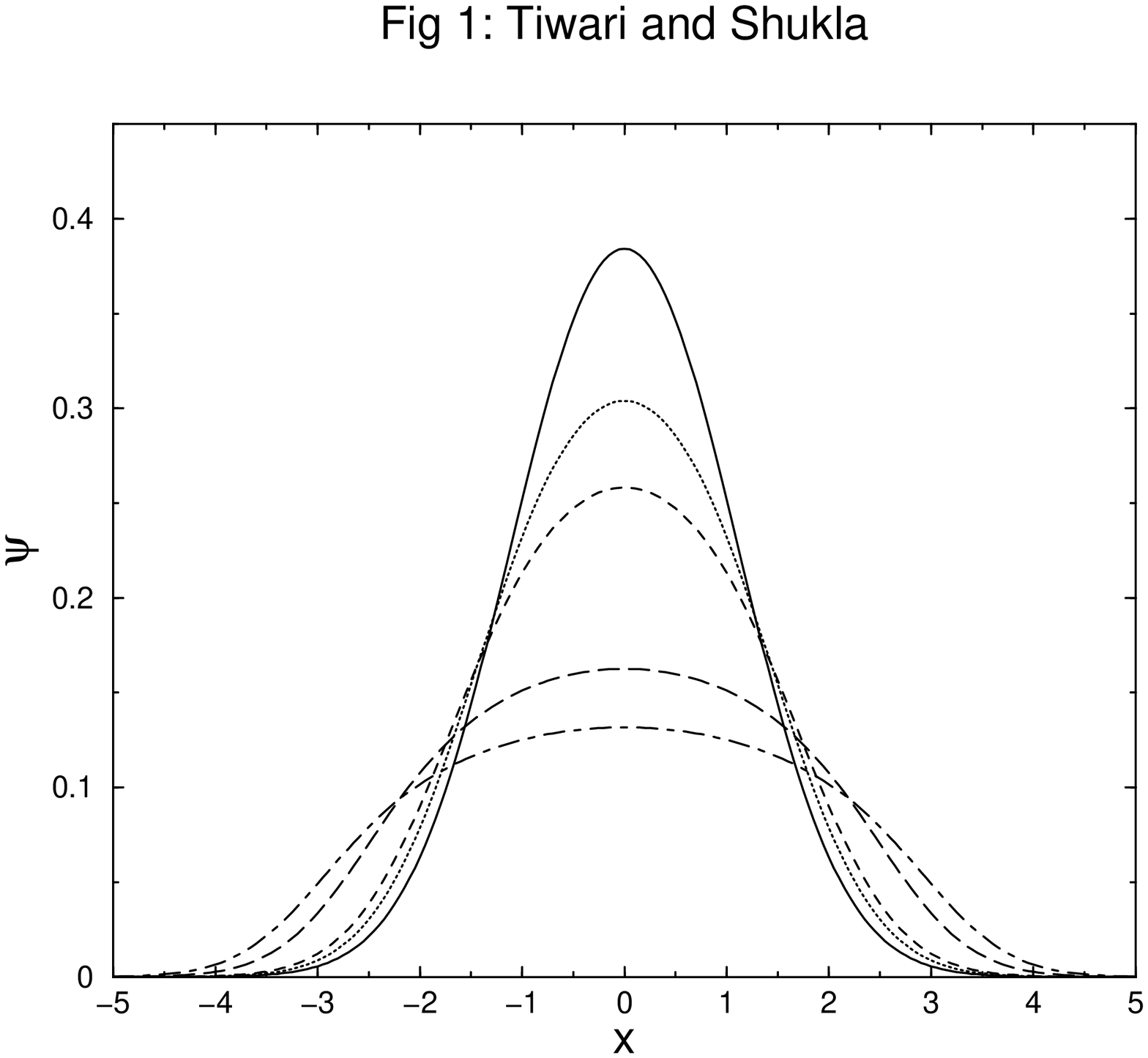}

\label{fig-cond}
\end{figure}

Finally, we present the plots of the condensates in a spherical trap,
for increasing values of $N$, in Fig. \ref{fig-cond}. As expected,
the calculations predict a depletion of central condensate density,
and corresponding delocalization of the condensate, with increasing
$N$. The results presented are in excellent agreement with similar
results presented by various other authors\cite{dalfovo,bao1}. Moreover,
if we compare the value of the condensate at the center of the trap
($|\psi(0,0,0)|$) for the isotropic trap with the published results
of Bao and Tang\cite{bao1}, we again obtain excellent agreement for
all values of $N$.

\subsection{Anharmonic Potentials}

\label{sub-anharmonic}
\begin{figure}

\caption{Influence of various types of anharmonicities on condensates in cylindrical
traps with $N=1000,$ and scattering length corresponding to the JILA
parameters\cite{jila}. The plots correspond to: (a) the Henon-Heiles
potential, and (b) the Fourleg potential. In each graph, solid, dotted,
and dashed lines represent values of anharmonicity parameter (see
text) $\alpha=0.0,\;0.05,$ and 0.15, respectively. The $y$ coordinate
is measured in the units of $a_{x}$.}

\includegraphics[%
  scale=0.6,
  angle=-90]{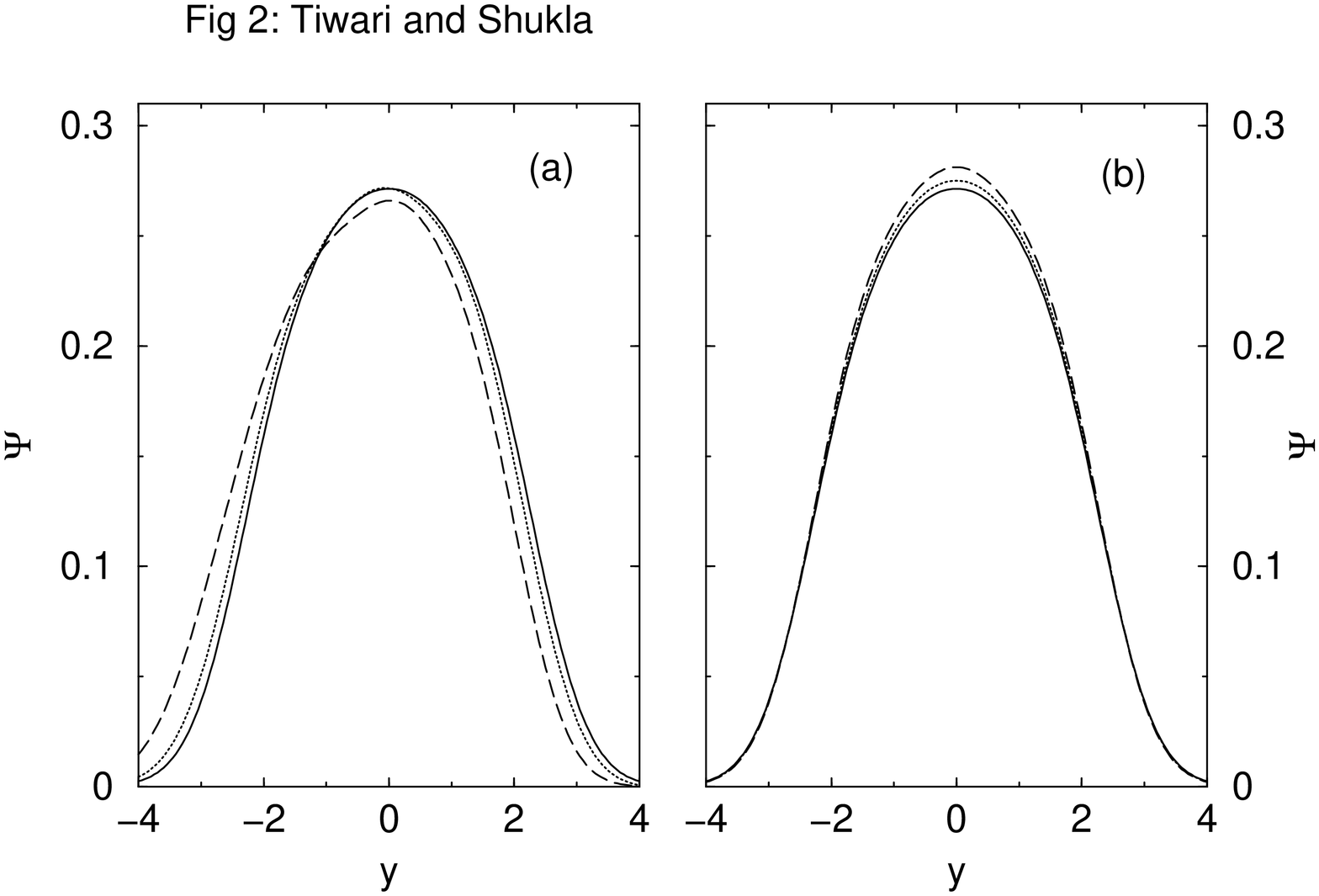}

\label{fig-anh}
\end{figure}

Recently, several studies have appeared in
the literature studying the influence of trap anharmonicities on the
condensates, in light of rotating condensates, and the resultant vortex
structure\cite{anharmonic}. However, we approach the influence of
trap anharmonicity from a different perspective, namely that of quantum
chaos. Therefore, the anharmonicities considered here are in the absence
of any rotation, and the aim is to study their influence on the ground
and the excited states of the condensate. We assume the unperturbed
harmonic trap to be the cylindrical one corresponding to the JILA
parameters\cite{jila}, and consider two types of anharmonic perturbations
in the $x-y$ plane: (a) the Henon-Heiles potential with $V^{anh}(x,y)=\alpha(x^{2}y-\frac{1}{3}y^{3})$,
and (b) the Fourleg potential $V^{anh}(x,y)=\alpha x^{2}y^{2}$, where
$\alpha$ is the anharmonicity parameter. Note that the Henon-Heiles
potential reduces the circular symmetry of the cylindrical trap in
the $x-y$ plane to the triangular one (symmetry group $C_{3v}$),
and the fourleg potential reduces the symmetry to that of a square
(group $C_{4v}$). In case of Henon-Heiles potential the inversion
symmetry of the cylindrical trap is also destroyed, while for the
fourleg potential, it is still preserved. The Henon-Heiles potential
introduces deconfinement in the trap, the Fourleg potential, on the
other hand, strengthens the confinement of the original trap. Both
these potentials are known to lead to chaotic behavior for higher
energy states, both at the classical and quantum-mechanical levels
of theories\cite{henon,fourleg}. In a separate work communicated
elsewhere, we have examined the excited states of condensates under
the influence of these potentials, in order to analyze the signatures
of chaotic behavior. In the present work, however, we intend only
to demonstrate the capabilities of our program as far as the anharmonicity
is concerned, and restrict ourselves only to the ground states of
the condensates in presence of these potentials. Results of our calculations
on the chemical potentials of condensates in a cylindrical trap corresponding
to the JILA experiment\cite{jila}, and $N=1000$, are presented in
table \ref{tab-mu-anh} as a function of anharmonicity $\alpha$.
Corresponding plots of the condensate along the $y$ axis are presented
in Fig. \ref{fig-anh}. As far as the influence of anharmonicity on
the chemical potential is concerned, from table \ref{tab-mu-anh}
we conclude that, for a given value of $N$, for increasing $\alpha$,
it increases for the Henon-Heiles potential, and decreases for the
fourleg potential. Similarly, upon examining the Fig. \ref{fig-anh},
we conclude that for the Henon-Heiles potential, the central condensate
density gets depleted with increasing $\alpha$, while the behavior
in case of the fourleg potential is just the opposite. Additionally,
the fact that the inversion symmetry is broken in case of Henon-Heiles
potential, is obvious from the asymmetry of corresponding condensate
plots.

\subsection{Entropy Calculations}

\label{sub-entropy}
\begin{figure}

\caption{Plots of Shannon entropy of condensates in an isotropic trap (solid
line) and cylindrical trap (dashed line), as a function of the dimensionless
strength parameter $\kappa=\frac{4\pi aN}{a_{x}}$.}

\includegraphics[%
  scale=0.8,
  angle=-90]{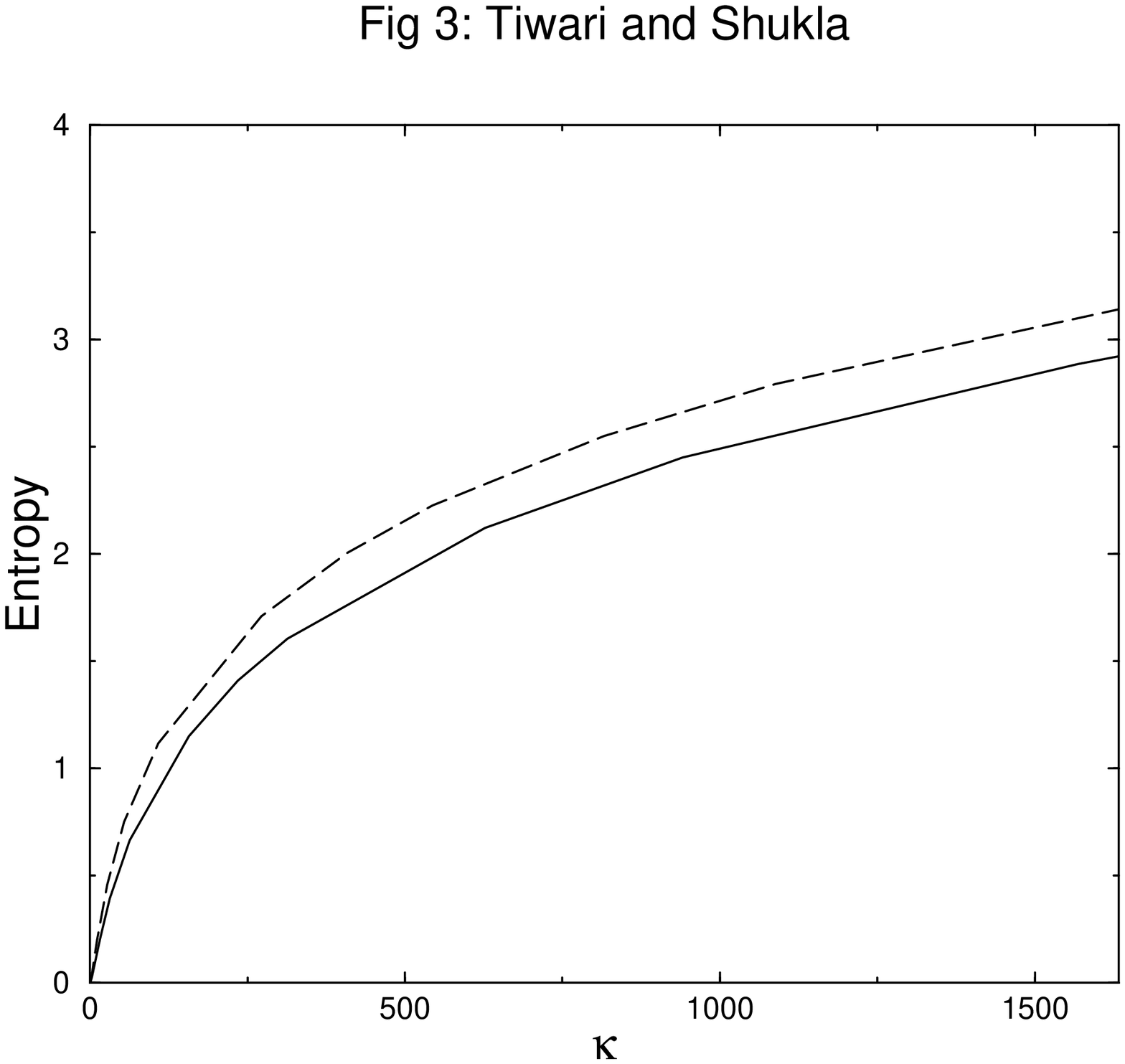}\label{fig-entropy}
\end{figure}

Here we discuss the Shannon entropy of condensates
in isotropic and cylindrical traps, as a function of the dimensionless
strength parameter $\kappa=\frac{4\pi aN}{a_{x}}$. Since the scattering
length in most of the traps is fixed, for such cases the change in
$\kappa$ can be construed as due to changes in $N$. In Fig. \ref{fig-entropy}
we present the plots of Shannon entropy versus $\kappa$ plots for
the condensates trapped both in isotropic, as well as cylindrical
traps. Although a detailed analysis of the Shannon entropy of condensates
is being presented elsewhere, we make a couple of important observations:
(i) For both types of traps the entropy increases as a function of
$\kappa$. Initially, the rate of increase is quite high, but for
larger values of $\kappa$, it settles down to a much lower value.
(ii) For a given nonzero value of $\kappa$, the entropy of a condensate
in a cylindrical trap is always larger than that of a condensate in
a spherical trap. In other words, the trap anisotropy appears to increase
the Shannon entropy of the system.

\section{Conclusions}

\label{sec-conclusions}In this paper we have reported a Fortran 90
implementation of a harmonic oscillator basis set based approach towards
obtaining the numerical solutions of time independent GPE. We have
presented applications of our program to a variety of situations including
anharmonic potentials, and in calculations of the Shannon entropy
of the condensate. We also compared the results obtained from our
program to those obtained by other authors, and found near-perfect
agreement. Therefore, we encourage the users to apply our program
to a variety of situations, and contact us in case they encounter
errors. We have extensive plans for further development of our program.
Some of the possible directions are: (a) extension of our approach
to time-dependent GPE, allowing one to deal with condensate dynamics,
(b) taking condensate rotation into account, allowing one to study
the vortex phenomena, and (c) dealing with condensates with nonzero
spins, i.e., the so-called spinor condensates\cite{spinor}. Work
along these lines is presently in progress in our group, and, upon
completion, will be reported in future publications.

\begin{appendix}

\section{Appendix}

Here our aim is to compute the two-particle integral of Eq. (\ref{eq-jmat})
defined as\begin{equation}
J_{n_{i}n_{j}n_{k}n_{l}}=\int_{-\infty}^{\infty}d\xi\phi_{n_{l}}(\xi)\phi_{n_{k}}(\xi)\phi_{n_{j}}(\xi)\phi_{n_{i}}(\xi),\label{eq-jdef}\end{equation}
where, in terms of the dimensionless coordinates $\xi$, the single
particle wave function $\phi_{n_{i}}(\xi)$ is given by\begin{equation}
\phi_{n_{i}}(\xi)=\frac{1}{\sqrt{\sqrt{\pi}2^{n_{i}}n_{i}!}}H_{n_{i}}(\xi)e^{-\frac{\xi^{2}}{2}}.\label{eq-phin}\end{equation}

Substituting Eq.\ref{eq-phin} in Eq. (\ref{eq-jmat}), we get 

\begin{equation}
J_{n_{i}n_{j}n_{k}n_{l}}=\frac{1}{\pi\sqrt{2^{n_{i}+n_{j}+n_{k}+n_{l}}n_{i}!n_{j}!n_{k}!n_{l}!}}I_{n_{i}n_{j}n_{k}n_{l}},\label{eq-JI}\end{equation}
 where \begin{equation}
I_{n_{i}n_{j}n_{k}n_{l}}=\int_{-\infty}^{\infty}e^{-2\xi^{2}}H_{n_{i}}(\xi)H_{n_{j}}(\xi)H_{n_{k}}(\xi)H_{n_{l}}(\xi)d\xi.\label{eq-Iint}\end{equation}

Since Hermite polynomials have a definite parity, the integral $I_{n_{i}n_{j}n_{k}n_{l}}$
will be nonvanishing only if the sum $n_{i}+n_{j}+n_{k}+n_{l}$ is
an even number. Now we will use a standard result for the product
of two Hermite Polynomials,\begin{equation}
H_{m}(\xi)H_{n}(\xi)=\sum_{k=0}^{\min\{ m,n\}}2^{k}k!\left(\begin{array}{c}
m\\
k\end{array}\right)\left(\begin{array}{c}
n\\
k\end{array}\right)H_{m+n-2k}(\xi),\label{eq-hermprod}\end{equation}

where $\left(\begin{array}{c}
m\\
k\end{array}\right)$ etc. are the binomial coefficients. Upon substituting Eq. \ref{eq-hermprod}
in Eq. \ref{eq-Iint}, we obtain\begin{equation}
I_{m,n,q,r}=\sum_{k=0}^{\min\{ m,n\}}\sum_{l=0}^{\min\{ q,r\}}2^{k+l}k!l!\left(\begin{array}{c}
m\\
k\end{array}\right)\left(\begin{array}{c}
n\\
k\end{array}\right)\left(\begin{array}{c}
q\\
l\end{array}\right)\left(\begin{array}{c}
r\\
l\end{array}\right)\int_{-\infty}^{\infty}e^{-2\xi^{2}}H_{m+n-2k}(\xi)H_{q+r-2l}(\xi)d\xi.\label{eq-Iint2}\end{equation}

In order to perform the integral above, we recall the result derived
by Busbridge\cite{busbridge}

\begin{equation}
\int_{-\infty}^{\infty}e^{-2\xi^{2}}H_{m}(\xi)H_{n}(\xi)d\xi=(-1)^{\frac{m-n}{2}}2^{\frac{m+n-1}{2}}\Gamma(\frac{m+n+1}{2}),\label{eq-busbridge}\end{equation}
 for $m+n=\textrm{even}$. Substituting this we get \begin{equation}
\begin{split}I_{m,n,q,r}= & (-1)^{\frac{m+n-p-q}{2}}2^{\frac{m+n+p+q-1}{2}}K_{m,n,q,r},\\
\end{split}
\label{eq-Iintf}\end{equation}
 where \begin{equation}
K_{m,n,q,r}=\sum_{k=0}^{\min\{ m,n\}}\sum_{l=0}^{\min\{ q,r\}}(-1)^{l-k}k!l!\left(\begin{array}{c}
m\\
k\end{array}\right)\left(\begin{array}{c}
n\\
k\end{array}\right)\left(\begin{array}{c}
q\\
l\end{array}\right)\left(\begin{array}{c}
r\\
l\end{array}\right)\Gamma(\frac{m+n+q+r-2k-2l}{2}+\frac{1}{2}).\label{eq-kmnqr}\end{equation}
 Upon substituting Eqs. (\ref{eq-Iintf}) and (\ref{eq-kmnqr}) in
Eq. (\ref{eq-JI}), we get\begin{equation}
J_{m,n,q,r}=(-1)^{\frac{m+n-q-r}{2}}\frac{1}{\pi\sqrt{2(m!n!q!r!)}}K_{m,n,q,r}.\label{eq-j1}\end{equation}

On using the expression\begin{equation}
\Gamma(\frac{2n+1}{2})=\frac{2n!}{2^{2n}n!}\sqrt{\pi},\label{eq-gamma2}\end{equation}

and setting $m+n+q+r=2t$, we have \begin{equation}
\Gamma(\frac{m+n+q+r-2k-2l}{2}+\frac{1}{2})=\frac{(2t-2k-2l)!\sqrt{\pi}}{2^{2t-2k-2l}(t-k-l)!}.\label{eq-gamma3}\end{equation}
 Upon substituting Eq. (\ref{eq-gamma3}) and the values of binomial
coefficients in Eq. (\ref{eq-kmnqr}), we have \begin{equation}
\begin{split}K_{m,n,q,r} & =\frac{\sqrt{\pi}}{2^{m+n+q+r}}m!n!q!r!\times\\
 & \sum_{k,l}\frac{(-1)^{l-k}2^{2(k+l)}(2t-2k-2l)!}{(m-k)!(n-k)!(q-l)!(r-l)!(k!)(l!)(t-k-l)!},\end{split}
\label{eq-kfinal}\end{equation}
 which, upon substitution in Eq. (\ref{eq-j1}), leads to the final
expression\begin{equation}
J_{m,n,q,r}=\frac{(-1)^{\frac{m+n-q-r}{2}}}{2^{m+n+q+r}}\sqrt{\frac{m!n!q!r!}{2\pi}}L_{m,n,q,r},\label{eq-jfinal}\end{equation}
 where \begin{equation}
L_{m,n,q,r}=\sum_{k=0}^{\min\{ m,n\}}\sum_{l=0}^{\min\{ q,r\}}\frac{(-1)^{l-k}2^{2k+2l}(2t-2k-2l)!}{(m-k)!(n-k)!(q-l)!(r-l)!(k!)(l!)(t-k-l)!}.\label{eq-lfinal}\end{equation}

Expressions of Eqs. (\ref{eq-jfinal}) and (\ref{eq-lfinal}) have
been used in the function JINTT, which is called via subroutine JMNPQ\_CAL,
to compute these two-body integrals. We would like to emphasize that
the series of Eq. (\ref{eq-lfinal}) has terms with alternating signs,
and, therefore, is potentially unstable for large values of $m,\: n,\: q,$
and $r$. Thus, it is crucial to use high arithmetic precision while
summing the series. With the usual double-precision arithmetic (REAL{*}8
variables), we found that the the series was unstable for values of
$m,\: n,\: q,\: r$ larger than 16. To circumvent these problems,
we used quadruple precision (REAL{*}16 variables) in function JINTT
to sum the series. Once the summation is performed, the results are
converted into the double-precision format. We believe that this approach
has made the two-particle integral calculation process very robust,
and accurate. 

\end{appendix}

\begin{table}

\caption{Convergence on the chemical potential and the mixing entropy of the
condensate in an isotropic trap with scattering length $a=2.4964249\times10^{-3}a_{x}$
and number of bosons $N=1500$, with respect to the basis set size.
$nmax$ is the maximum value of the quantum number of the SHO basis
function in a given direction, $N_{basis}$ is the total number of
basis functions corresponding to a given value of $nmax$, and $N_{iter}$
represents the total number of SCF iterations needed to achieve convergence
on the condensate energy per particle. In all the calculations iterative
diagonalization method, along with Fock matrix mixing with $xmax=0.6$,
was used. The SCF convergence threshold was $1.0\times10^{-7}$. }

\begin{tabular}{|c|c|c|c|c|}
\hline 
$nmax$&
$N_{basis}$&
$N_{iter}$&
Chemical Potential&
Entropy\tabularnewline
\hline
\hline 
2&
4&
19&
2.939116&
0.5083669\tabularnewline
\hline 
4&
10&
11&
2.915046&
0.5387074\tabularnewline
\hline 
6&
20&
14&
2.911181&
0.5396975\tabularnewline
\hline 
8&
35&
14&
2.911278&
0.539088\tabularnewline
\hline 
10&
56&
15&
2.911375&
0.5392423\tabularnewline
\hline 
12&
84&
15&
2.911346&
0.5392770\tabularnewline
\hline 
14&
120&
15&
2.911337&
0.5392822\tabularnewline
\hline 
16&
165&
13&
2.911337&
0.5392797\tabularnewline
\hline
\end{tabular}\label{tab-conv-iso}
\end{table}

\begin{table}

\caption{Convergence on the chemical potential and the mixing entropy of the
condensate in a cylindrical trap with trap parameters corresponding
to the JILA experiment\cite{jila}, and the number of bosons $N=1500$,
with respect to the basis set size. $nxmax$ is the maximum value
of the quantum number of the SHO basis function in $x$- and $y-$direction,
$nzmax$ is the same number corresponding to the $z$-direction. Rest
of the quantities have the same meaning as explained in the caption
of table \ref{tab-conv-iso}. In all the calculations iterative diagonalization,
along with Fock matrix mixing with $xmax=0.3$, was used. The SCF
convergence threshold was $1.0\times10^{-8}$.\label{tab-conv-cyl}}

\begin{tabular}{|c|c|c|c|c|c|}
\hline 
$nxmax$&
$nzmax$&
$N_{basis}$&
$N_{iter}$&
Chemical Potential&
Entropy\tabularnewline
\hline
\hline 
4&
0&
6&
39&
5.786323&
0.9387124\tabularnewline
\hline 
4 &
2&
12&
19&
5.421930&
0.9383227\tabularnewline
\hline 
4&
4&
18&
27&
5.423981&
0.9380013\tabularnewline
\hline 
6&
0&
10&
31&
5.737602&
0.9770727\tabularnewline
\hline 
6&
2&
20&
19&
 5.405221&
0.9549194\tabularnewline
\hline 
6&
4&
30&
20&
5.405440&
0.9545441\tabularnewline
\hline 
6&
6&
40&
20&
5.404970&
0.9545920\tabularnewline
\hline 
8&
0&
15&
21&
5.737074&
0.9752491\tabularnewline
\hline 
8&
2&
30&
20&
5.404605&
0.9546530\tabularnewline
\hline 
8&
4&
45&
21&
5.404691&
0.9542773\tabularnewline
\hline 
8&
6&
60&
21&
5.404218&
0.9543291\tabularnewline
\hline 
8&
8&
75&
21&
5.404147&
0.9543239\tabularnewline
\hline 
10&
0&
21&
21&
5.736935&
 0.9755365\tabularnewline
\hline 
10&
2&
42&
20&
5.404741&
0.9544814\tabularnewline
\hline 
10&
4&
63&
20&
5.404540&
0.9541044\tabularnewline
\hline 
10&
6&
84&
20&
5.404066&
0.9541561\tabularnewline
\hline 
10&
8&
105&
18&
5.403994&
0.9541508\tabularnewline
\hline 
10&
10&
126&
18&
5.403991&
0.9541477\tabularnewline
\hline 
12&
0&
28&
21&
5.736763&
0.9756539\tabularnewline
\hline 
12&
2&
56&
20&
5.404636&
0.9545789\tabularnewline
\hline 
12&
4&
84&
20&
5.404436&
0.9542004\tabularnewline
\hline 
12&
6&
112&
20&
5.403962&
0.9542517\tabularnewline
\hline 
12&
8&
140&
20&
5.403891&
0.9542463\tabularnewline
\hline 
12&
10&
168&
20&
5.403888&
0.9542432\tabularnewline
\hline 
12&
12&
196&
18&
5.403888&
0.9542428\tabularnewline
\hline
\end{tabular}
\end{table}

\begin{table}

\caption{Comparison of the chemical potentials ($\mu$) obtained using the
iterative diagonalization (ID) technique, and the steepest-descent
(SD) technique\cite{dalfovo}, for the condensates in a cylindrical
trap with trap parameters corresponding to the JILA experiment\cite{jila},
and a given number of bosons ($N$). Quantities $N_{iter}$, $nxmax$,
and $nzmax$ have the same meaning as in the caption of table \ref{tab-conv-cyl},
and $\mu$ is expressed in the units of $\hbar\omega_{x}$. In the
ID based calculations for $N\leq2000$, SHO ground state solution
was used to start the iterations, while for larger values of $N$
the iterations were started using the Thomas-Fermi approximation.
In all the cases corresponding to the ID method, Fock matrix mixing
was used, with $0.05\leq xmax\leq0.3$. In the SD based calculations,
the iterations were started using the Thomas-Fermi approximation,
with the size of the time step being $0.02$ units.\label{tab-comp-method}}

\begin{tabular}{|c|c|c|c|c|c|c|}
\hline 
$N$&
$nxmax$&
$nzmax$&
$N_{iter}$(ID)&
$N_{iter}$(SD)&
$\mu$(ID)&
$\mu$(SD)\tabularnewline
\hline
\hline 
500&
8&
6&
16&
62&
3.938611&
3.938865\tabularnewline
\hline 
1000&
8&
6&
14&
88&
4.770707&
4.772055\tabularnewline
\hline 
1500&
8&
6&
21&
95&
5.404218&
5.405491\tabularnewline
\hline 
2000&
12&
8&
23&
104&
5.931870&
5.932878\tabularnewline
\hline 
10000&
14&
8&
86&
99&
10.505267&
10.505124\tabularnewline
\hline 
15000&
14&
8&
105&
129&
12.239700&
12.239465\tabularnewline
\hline 
20000&
14&
8&
104&
109&
13.665923&
13.665686\tabularnewline
\hline
\end{tabular}
\end{table}

\begin{table}

\caption{Comparison of the chemical potentials (in the units of $\hbar\omega$)
obtained from our program, and those reported by Bao and Tang\cite{bao1},
for an isotropic trap, with increasing values of interaction parameter
$\kappa$. The negative value of $\kappa$ implies attractive inter-particle
interactions. For the value of scattering length stated in table \ref{tab-conv-iso},
$\kappa=3137.1$ corresponds to $N=1\times10^{5}$ bosons. Symbols
$nmax$ and $N_{iter}$ have the same meaning as in the previous tables.
For the last two calculations, SD method with a time step of 0.02
units, and Thomas-Fermi initial guess were employed. Our chemical
potentials have been truncated to as many decimal places as reported
by Bao and Tang\cite{bao1}. }

\begin{tabular}{|c|c|c|c|c|}
\hline 
$\kappa$&
$nmax$&
$N_{basis}$&
$\mu$(This work)&
$\mu$(Ref.\cite{bao1})\tabularnewline
\hline
\hline 
-3.1371&
14&
120&
1.2652&
1.2652\tabularnewline
\hline 
3.1371&
14&
120&
1.6774&
1.6774\tabularnewline
\hline 
12.5484&
14&
120&
2.0650&
2.0650\tabularnewline
\hline 
31.371&
14&
120&
2.5861&
2.5861\tabularnewline
\hline 
125.484&
14&
120&
4.0141&
4.0141\tabularnewline
\hline 
627.42&
16&
165&
7.2485&
7.2484\tabularnewline
\hline 
3137.1&
16&
165&
13.553&
13.553\tabularnewline
\hline
\end{tabular}\label{tab-bao}
\end{table}

\begin{table}

\caption{Comparison of the chemical potentials (in the units of $\hbar\omega_{x}$)
obtained from our program, and those reported by Dalfovo and Stringari\cite{dalfovo},
for a cylindrical trap corresponding to the JILA parameters\cite{jila},
with increasing number $N$ of bosons. Symbols $nxmax$, $nzmax$,
and $N_{iter}$ have the same meaning as in the previous tables. Calculations
for $N\geq10000$ were performed by the SD method using Thomas-Fermi
starting orbitals, a time-step of $0.02$ units, and a convergence
threshold of $1.0\times10^{-7}$. We have truncated our chemical potentials
to as many decimal places as reported by Dalfovo and Stringari\cite{dalfovo}.}

\begin{tabular}{|c|c|c|c|c|c|}
\hline 
$N$&
$nxmax$&
$nzmax$&
$N_{basis}$&
$\mu$(This work)&
$\mu$(Ref.\cite{dalfovo})\tabularnewline
\hline
\hline 
100&
8&
6&
60&
2.88&
2.88\tabularnewline
\hline 
200&
8&
6&
60&
3.21&
3.21\tabularnewline
\hline 
500&
8&
6&
60&
3.94&
3.94\tabularnewline
\hline 
1000&
8&
6&
60&
4.77&
4.77\tabularnewline
\hline 
2000&
8&
6&
60&
5.93&
5.93\tabularnewline
\hline
5000&
10&
8&
105&
8.14&
8.14\tabularnewline
\hline
10000&
10&
8&
105&
10.5&
10.5\tabularnewline
\hline
15000&
14&
8&
180&
12.2&
12.2\tabularnewline
\hline
20000&
14&
8&
180&
13.7&
13.7\tabularnewline
\hline
\end{tabular}

\label{tab-dalfovo}
\end{table}

\begin{table}

\caption{Influence of trap anharmonicities on the chemical potential. The
table below presents results for the Henon-Heiles, and the fourleg
potentials, for cylindrical trap corresponding to JILA parameters\cite{jila},
with $N=1000$.}

\begin{tabular}{|c|c|c|}
\hline 
$\alpha$&
$\mu$(Henon-Heiles)&
$\mu$(Fourleg)\tabularnewline
\hline
\hline 
0.00&
4.7712&
4.7712\tabularnewline
\hline 
0.03&
4.7662&
4.8261\tabularnewline
\hline 
0.06&
4.7497&
4.8752\tabularnewline
\hline 
0.09&
4.7207&
 4.9202\tabularnewline
\hline 
0.12&
4.6764&
4.9619\tabularnewline
\hline 
0.15&
4.6131&
5.0009\tabularnewline
\hline
\end{tabular}

\label{tab-mu-anh}
\end{table}

\end{document}